# PISCES-RF: a liquid-cooled high-power steady-state helicon plasma device

Saikat Chakraborty Thakur[a,c], Michael J. Simmonds[a], Juan F. Caneses[b], Fengjen Chang[a], Eric M. Hollmann[a], Russell P. Doerner[a], Richard Goulding[b], Arnold Lumsdaine[b], Juergen Rapp[b] and George R. Tynan[a]

*[a] University of California, San Diego, San Diego, CA, USA*
*[b] Oak Ridge National Laboratory, Oak Ridge, TN, USA*
*[c] Department of Physics, Auburn University, Auburn, AL 36849*

Radio-frequency (RF) driven helicon plasma sources can produce relatively high-density plasmas ($n > 10^{19}$ m$^{-3}$) at relatively moderate powers (< 2 kW) in argon. However, to produce similar high-density plasmas for fusion relevant gases such as hydrogen, deuterium and helium, much higher RF powers are needed. For very high RF powers, thermal issues of the RF-transparent dielectric window, used in the RF source design, limit the plasma operation timescales. To mitigate this constraint, we have designed, built and tested a novel helicon plasma source assembly with a fully liquid-cooled RF-transparent window which allows steady state operations at high power (up to 20 kW) and successfully produces high-density plasma with both argon and hydrogen. De-ionized (DI) water, flowing between two concentric dielectric RF windows, is used as the coolant. We show that a full azimuthal blanket of DI water does not prevent high-density plasma production. From calorimetry on the DI water, we measure the net heat removed by the coolant at steady state conditions. Using infra-red (IR) imaging, we calculate the constant plasma heat deposition and measure the final steady state temperature distribution patterns on the inner surface of the ceramic layer. The heat deposition pattern follows the helical shape of the antenna. We also show the consistency between the heat absorbed by the DI water, as measured by calorimetry, and the total heat due to the combined effect of the plasma heating and the absorbed RF. These results are being used to answer critical engineering questions for the 200 kW RF device (MPEX: Materials Plasma Exposure eXperiment) being designed at the Oak Ridge National Laboratory (ORNL) as a next generation plasma material interaction (PMI) device.

## 1. Introduction

The plasma facing components in the divertor regions of toroidal magnetically confined thermonuclear fusion reactors such as ITER, DEMO or any other future burning plasma devices will face extreme heat and plasma fluxes [please see Ref. 1 and the references within for a detailed review on this subject]. Understanding and mitigating the unwanted effects of large plasma fluxes and fluences on plasma facing components is key to the successful and long-lasting operation of future fusion reactors. In this regard, plasma material testing facilities serve a crucial role in thermonuclear fusion research [Ref. 2 and the references within]. However, the ability to examine plasma materials interactions (PMI) and the evolution of the plasma-material interface at reactor-relevant divertor fluences is not available in current (or planned) toroidal devices, mostly due to low duty cycle, lack of proper diagnostic access and lack of controlled experiments [3, 4]. Linear plasma facilities simulating plasma facing conditions have filled in this gap with the capability to achieve high fluences to serve as PMI testbeds [Section 3 of Ref. 2] with state-of-the-art diagnostic capabilities, easy access and a well-controlled plasma environment. In these devices, it is desired to achieve Hydrogen (H), Deuterium (D) and Helium (He) plasmas with particle fluxes $> 10^{22}$ m$^{-2}$sec$^{-1}$ and diameters of a few cm(s). Please see the Section 3 of Ref. 2 for a more comprehensive summary of most of the current linear plasma devices extensively used for PMI studies.

The standard mechanism of plasma production in most of the currently used linear PMI devices, that can produce high density ($10^{18}$ m$^{-3}$ – $10^{19}$ m$^{-3}$), steady state plasmas, is to use heated lanthanum hexaboride (LaB$_6$) [5] cathode sources or arc sources. Depending on the details of the design of the arc sources, there are typically either reflex arc sources [6 – 9] compatible with magnetic fields of up to ~ 0.1T or high pressure, pulsed, cascaded arc sources [10, 11], which works better at magnetic fields of ~ 1T. These have been the standard method of producing relatively high-density plasmas for PMI studies over the last few decades. However, the design of these sources require that they are in direct contact with the high-density plasma, which can lead to several additional issues such as introduction of impurities in the plasma, finite lifetime of the cathode and difficulty in reproducibility after long machine maintenance breaks.

On the other hand, radio frequency (RF) based plasma sources have been used in the basic plasma physics community for the last few decades to produce similar densities in argon plasma [12 – 15]. Depending on the external magnetic field and power, these RF devices can operate in the capacitive (E), inductive (H) or the helicon mode (W). In the helicon mode (W) of operation, RF sources can produce peak plasma densities of ~ 10$^{19}$ m$^{-3}$ in argon for a few kW (kilowatts) of RF powers in plasma chambers that are several meters long [16, 17]. Helicon plasma sources have the additional benefit of having the antenna mounted outside of the vacuum chamber which minimizes impurities in the plasma. Moreover, the axial access to the vacuum chamber in the source end of the device is unobstructed by the source itself (unlike cathodes or arc discharge sources) and hence can be directly used for diagnostic access and laser heating. Also, RF devices are very repeatable even after long device downtimes. Thus, helicon plasma sources have the

*Author's email: saikat@eng.ucsd.edu*

potential to be used in PMI devices if similar densities can be achieved in steady state in fusion relevant gases. However, plasmas with lighter ions have a higher thermal velocity and would need more RF power to sustain similar high densities. Hence, in the past few years, there has been a sustained research endeavor to build RF sources for PMI studies using much higher powers [18 – 22].

Helicon plasma sources require an insulating sleeve (dielectric window) for the RF to penetrate and produce high-density plasmas. For very high RF powers, thermal issues on the cylindrical "RF – transparent" dielectric window limit the plasma operation timescales. The current high-power helicon plasma sources designed for PMI studies are hence all operated in the pulsed mode with the plasma being ON for only up to few hundreds of milliseconds [20, 21]. To increase the net fluence on to the material target to values compatible with fusion reactor plasmas, it is desirable to accomplish steady state operations with low mass, high density plasmas. To allow steady state operations, we have to mitigate the thermal constraints, while maintaining the mechanical integrity of the RF source. Hence, we have designed, built and tested a novel liquid-cooled RF window which allows steady state operations at high power (up to 20 kW) with both argon and hydrogen plasmas. This is the first example reported in the literature where the RF dielectric window is fully immersed in a completely azimuthal blanket of a liquid coolant, for maximum uniformity in cooling.

As the only other example of a water-cooled dielectric window for a helicon source, in the experiment RAID [22], cooling channels have been drilled into the window, at 8 azimuthal locations, to keep the ceramic relatively cold for plasmas up to 5 kW of RF power. At these powers, hydrogen plasma with densities $< 10^{18}$ m$^{-3}$ were formed. But for steady state operations in PMI relevant high density ($> 10^{19}$ m$^{-3}$) hydrogen or deuterium plasma, more uniform cooling is required, especially as new PMI devices, such as the Materials Plasma Exposure eXperiment (MPEX) at Oak Ridge National Laboratory (ORNL), are being designed for up to 200 kW of RF power at the plasma source [23, 24]. The design of the liquid cooled RF source assembly described here is motivated by stringent requirements of the MPEX device. The conceptual design and preliminary numerical analysis of the performance of a water-cooled ceramic window for the MPEX device has been mainly based on short pulse thermal data of an uncooled source, currently used in Proto-MPEX [25]. Further detailed numerical studies of MPEX source design and performance are based on the work presented in this paper [26].

In this paper, we investigate the performance of the DI water-cooled RF plasma source design and focus on the following research questions: (1) What fraction of the RF power is absorbed by the DI water? (2) Is the high-density helicon mode accessible with an azimuthally fully enclosed blanket of cooling DI water? (3) What are the heat and steady state temperature distributions on the inner plasma facing surface of the dielectric window? (4) Will the cooling system work as expected and remove the heat deposited on the ceramic by the plasma in real time to achieve thermally steady state conditions? (5) How do the important engineering parameters scale with increasing RF power?

## 2. Experimental set up

### 2.1 The plasma device

Fig. 1 shows the CAD (Computer-Aided Design) model of the experimental device, pointing out some of the important aspects of the experiment. During actual plasma operation, the source assembly (Helicon Antenna area in Fig. 1) is covered with RF shielding to prevent RF noise escaping into the room, which can disrupt or damage sensitive diagnostics. Also, not shown is the matching box that is connected to the leads of the antenna (which are shown as copper rods sticking out at an angle of ~ 45 degrees). The new RF source assembly with the DI water cooled RF transparent window (detailed description is given in the next section 2.2) was incorporated on to an existing linear magnetized plasma chamber, located in the PISCES (Plasma Interaction Surface Component Experimental Station) laboratory at the University of California at San Diego (UCSD). Since the primary goal of this new device would be to produce steady state plasmas at high densities to generate extremely high fluences for PMI studies, we shall henceforth call this machine PISCES-RF.

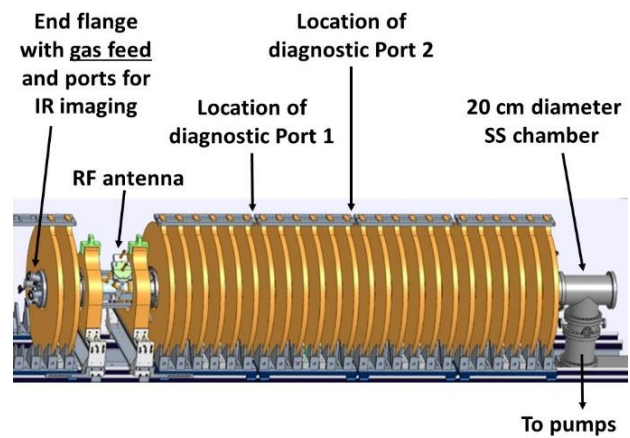

Fig. 1. CAD model of the PISCES-RF device.

The main cylindrical plasma chamber, along with the external magnets and pumping system belonged to the erstwhile Controlled Shear Decorrelation eXperiment (CSDX) [17, 27 – 38]. Previously in CSDX, argon plasmas with high density and relatively low temperatures ($n > 10^{19}$ m$^{-3}$, $T_{electron}$ ~ 5 ev and $T_{ion}$ < 1 eV [39]) were produced using an m = 1 helicon antenna, operating at 13.56 MHz at RF powers of ~ 1.5 – 2 kW. The current plasma device consists of a 3 m long, 20 cm diameter stainless steel (SS) chamber immersed in an external magnetic field that can go up to 0.240 T. In this work, the magnetic field used is almost uniform for most of the length of the machine, except at the location of the helicon source, which has a small ripple due to the extra separation of the two magnetic field coils around the RF antenna (please see Fig. 2). This extra space allowed us to safely install the high-power RF antenna assembly and build a grounded RF shield around it (please see Fig. 3).

This also ensured that the plasma following the magnetic field lines gets radially terminated at the center of the ceramic (as shown in Fig. 2), thus allowing us to test the worst possible condition for ceramic heating.

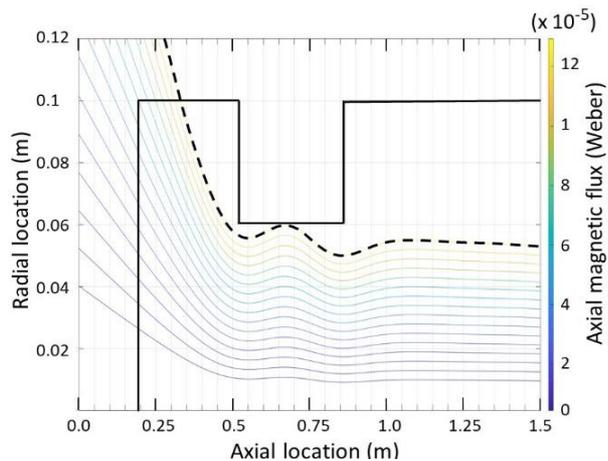

Fig. 2. The axial magnetic flux near the RF source end of the PISCES-RF device. The dark solid lines represent the plasma chamber (radius = 0.1 m). The dark line at ~ r = 0.06 m represents the location of the ceramic window, over which the RF antenna is wrapped. The dashed black line represents the magnetic flux line in direct contact with the ceramic.

A turbomolecular pump, placed at the downstream end of the chamber, backed by two other mechanical pumps allows the base pressure to be in the high $10^{-7}$ Torr. A manually controlled valve placed near the pumps allow us to change the neutral pressures in the device independently from the gas flow rate. Depending on the gas used, to achieve the helicon mode of operation (W), we typically need the neutral gas to flow at 20 – 100 sccm (standard cubic centimeters per minute) which gives us neutral pressures of 2 – 10 mTorr in the plasma chamber. In the current set up, we have added a specially designed end flange (please see Fig. 2 for the details) upstream of the RF source that allows upstream end gas injection. This gas injection configuration has shown to critically affect the neutral gas management in Proto-MPEX [40] and is also the choice of operation for MPEX. Hence, we incorporated this change on PISCES-RF.

**2.2 DI water cooled RF source assembly**

To operate the device with up to 20 kW of RF power, the helicon window must be actively liquid cooled. In our design, we used two concentric cylindrical dielectric tubes with De-ionized (DI) water flowing between them. This double walled RF transparent window material would require dielectrics with a very low loss tangent. Hence, we chose the inner window to be alumina ($Al_2O_3$), having a thickness of ~ 6.35 mm (0.25 inches), while the outer window is quartz with thickness of ~ 3 mm (0.117 inches). The annular water channel in between the two ceramics is ~ 3.35 mm (0.133 inches) wide. The inner diameter of the alumina cylinder is ~ 12 cm (4.75 inches). The RF transparent ceramic window assembly has an outer diameter of ~ 14.6 mm (5.75 inches). This assembly is attached to conflat flanges on the two ends so that the whole assembly is compatible to both the CSDX chamber at UCSD and the Proto-MPEX chamber at ORNL.

DI water is used as the coolant to reduce RF absorption by the coolant itself. We used a recirculating heat exchanger to cool the DI water and re-route it back to the source. Since DI water can be corrosive to copper or brass, we designed the whole recirculating chiller and the corresponding plumbing system using PVC piping and SS adapters. Note that the RF antenna itself is made from copper, and hence the cooling lines for the antenna itself are kept separate from the DI water coolant lines for the ceramic. Moreover, to reduce impurities in the plasma, a ceramic-to-metal joint is expected to be used in MPEX for the water-to-vacuum boundary. Hence, we have used a ceramic to metal sealing technology using titanium brazing, assembled for us by MPF Inc.

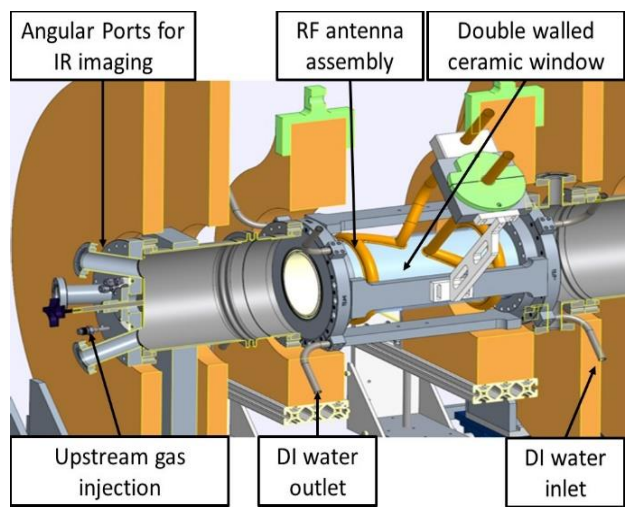

Fig. 3. CAD model showing the details of the new DI water cooled RF source assembly and the upstream end flange.

In Fig. 3, we show the main aspects of the new source assembly. The SS chambers on either side are shown in dark gray. The external magnets are shown in dark yellow. The ceramic RF transparent window is shown in light blue (with white inner ceramic), over which the water-cooled helicon antenna (shown in golden) is wrapped. DI water is introduced in between the two concentric ceramic layers through bores drilled into the SS flanges, at 4 symmetrically placed azimuthal locations. The upstream end flange has been specially designed with four angular ports for infrared (IR) camera access to get a full azimuthal view of the inner ceramic surface. The end flange is also fitted with ultra-high vacuum compatible valves for end gas injection. We also have a water-cooled end plate, which acts as a shutter (position controlled by the black knob at the right end in Fig. 3.) for the IR windows when the IR camera is not being used and also as an upstream dump plate for the plasma.

**3. Diagnostics**

3.1 Dielectric losses

In an RF source, we need a matching network for efficient coupling of the RF power from the power supply to the antenna. In the helicon plasma source design investigated in this paper, a layer of DI water fully separates the helicon antenna and the plasma volume. We are interested in the fraction of RF power absorbed in this fully enclosing layer and its impact on high-density plasma production.

We used a Vector Network Analyzer (VNA) to measure the RF losses due to the DI water coolant, by measuring the return loss of the RF matching circuit when attached to the antenna with and without the coolant in between the two ceramic layers. The return loss, which is given by

$$Return\ Loss\ [dB] = 10 \cdot \log_{10}(P_{reflected}/P_{forward}), \quad (1)$$

is measured as a function of the input frequency in the absence of plasma. At the resonant condition, the reflection is minimum, indicating maximum absorption of RF power by the loaded resonant circuit. The quality factor ($Q_L$) of a loaded resonant circuit is related to the width ($\Delta f$) of the return loss curve, which in turn is related to the total resistance of the resonant circuit ($R_T$), by the equation

$$Q_L = f_0/\Delta f = (\omega_0 L)/(2R_T). \quad (2)$$

Here $R_T = R_V + R_D + R_P$, where $R_V$ is the antenna vacuum resistance, $R_D$ is the resistance due to the dielectric window and $R_P$ is the plasma resistance, $f_0$ is the resonant frequency, $\omega_0$ is the angular frequency at resonance and $L$ is the inductance of the RF antenna. The width of the return loss curve can be calculated from the values of the input frequencies at the 3 dB points and the corresponding resistance is given by

$$R_T = \pi L \Delta f. \quad (3)$$

Thus, by measuring the 3-dB width of the return loss curve using the VNA in the absence of plasma ($R_P = 0$), with and without the coolant, we can calculate the difference in the RF loading ($R_T$), which gives an estimate of the RF losses due to the coolant.

To measure the RF losses in the presence of plasma, we used a Pearson transformer to directly measure the current through the antenna. From the measured forward power to the RF antenna, we can calculate the effective resistive loading. Again, we perform these experiments with and without the coolant to measure the difference in the resistive plasma loading ($R_T$). This method can also be used to measure the resistive vacuum loading ($R_V$).

### 3.2 Infra-red (IR) thermography

To measure the temperature of the plasma facing inner surface of the ceramic (alumina), we used an FLIR T420 camera operating at 30 Hz, with a spatial resolution of 240 x 320 pixels for infrared (IR) imaging. Due to mechanical constraints, the IR camera was placed at a shallow angle, as shown in Fig. 2 (the line to the center of the ceramic is ~ 17º with respect to the horizontal plane). This ensured that the IR view encompasses the full axial range of the inner surface of the ceramic. Four such symmetrically placed angular ports allowed us to obtain a full azimuthal view of the inner ceramic surface. From the IR camera location, the cylindrical ceramic surface looked like a part of a truncated cone, which was then mathematically mapped to a z – θ plane. Each individual azimuthal position of the IR camera recorded a little more than ± 90º azimuthal view of the inner surface of the cylindrical ceramic. With IR imaging data from four such locations, we mathematically reconstructed the full z – θ thermal maps of the cylindrical surface. We also found that the shallow angle of the viewport allowed the IR emission from the uncooled, hot chamber walls to reflect off the polished white ceramic surface back onto the IR camera. This gave a spurious contribution to the IR data, which had to be corrected for. We performed calibration by cooling the chamber walls using liquid nitrogen. The mathematical mapping algorithms and the methods used to take care of the spurious IR reflection, that could otherwise lead to erroneous interpretation, are being presented elsewhere [41].

### 3.3 Calorimetry

We also used two thermocouples to measure the inlet and the outlet temperatures of the coolant DI water, as a function of time. From the difference of these measurements, we could use calorimetry to calculate the net amount of heat that was being removed by the coolant. This also gave us the timescales of how fast the system reached steady state. This information turned out to be an important parameter in understanding and determining the steady state temperature from the IR camera analysis, that allowed us to get rid of the spurious contribution from the back reflection of the hot chamber during IR imaging.

### 3.4 Langmuir probes

Finally, we used two standard RF-compensated, voltage-swept Langmuir probes to measure the typical plasma parameters such as the plasma density, electron temperature and plasma potentials at two axial locations, 0.8 m and 1.5 m downstream of the RF antenna.

## 4. Experimental Results

### 4.1 Effect of DI water on RF absorption and high-density plasma production

The foremost concern regarding this liquid cooled RF antenna design was the fact that there is a fully azimuthal blanket of liquid under the RF antenna. This section explores the effect this has on RF absorption and high-density plasma production.

As explained in section 3, we used the VNA to measure the return loss of the matching circuit, loaded by the RF antenna, both with and without the DI water coolant. In Fig. 4, one example of such a return loss measurement is shown. Red circles are data with the coolant flowing, while the black squares are without any coolant in between the two ceramics, and this same scheme is used for Fig.(s) 4 – 8. From the return loss curves shown in Fig. 4, we calculated the widths using the upper and lower frequencies at the 3-dB cut off points and used the resonance frequency and equations (2) and (3) to calculate the quality factor ($Q_L$) and the vacuum resistance ($R_V$). In this case, when we do not have the DI water coolant in the RF source, $Q_L = 125$ and $R_V = 0.164\ \Omega$. When we introduce DI water coolant, we get $Q_L = 112$ and $R_V = 0.184\ \Omega$. Thus, the effect of introducing the DI water coolant was a reduction of the quality factor by 11% and increase in the vacuum loading resistance by ~ 0.02 $\Omega$, which is not very concerning. For many scans over multiple days, this factor ranged from 7% to 11%. We also notice that the resonance frequency shifts by ~ 0.08 MHz. This is expected, as introduction of the DI water coolant slightly changes the impedance of the RF antenna assembly. But this small change in the resonance frequency could be easily taken care of by the tuning capacitors in the matching circuit.

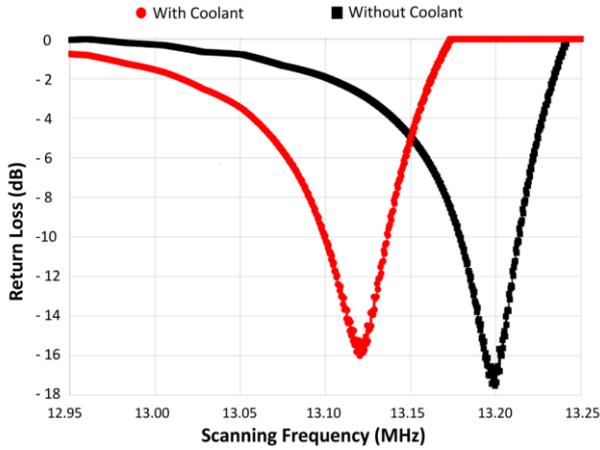

Fig. 4. Return loss, measured using a VNA, showing the RF absorption with (red circles) and without (black squares) the DI water coolant in the source.

In Fig(s). 5 and 6, we show the results of RF loading measurements using the Pearson transformer in the cases with no plasma (RF power was turned on, but the neutral gas was not injected into the chamber) and with plasma operation, respectively. For each chosen RF power input, we measured the corresponding current through the leads of the RF antenna and calculated the resistive loading. We show the normalized resistive loading with and without the DI water coolant. We found that the effect of introducing the DI water coolant was to effectively increase the resistive loading, which is consistent with what we found using the VNA. For vacuum conditions, the increase in the loading varied by 5% to 9% of the original antenna loading (without the DI water coolant).

As in most typical RF driven helicon sources, during operation, we observe several mode-jumps from the Capacitively Coupled Plasma (E mode: CCP) to Inductively Coupled Plasma (H mode: ICP) and finally the helicon mode (W). We can observe the effect of these mode jumps on the resistive loading of the matching circuit, even though these differential measurements can introduce a lot of uncertainty (as we are taking a small difference between two large numbers). These RF power values of the mode jumps, as seen in Fig. 6, are consistent with the typical visual characteristics of mode jumps in similar helicon devices [16, 17]. More importantly, we can clearly see the effect of the DI water coolant on the resistive loading. In our regime of interest (high power helicon mode of operation), the effect of DI water coolant on the loading is to increase it by 9 ± 1 %. Since this experiment required operating the source without the DI water coolant, we did not operate the machine with higher than 3 kW of RF power, for safety issues.

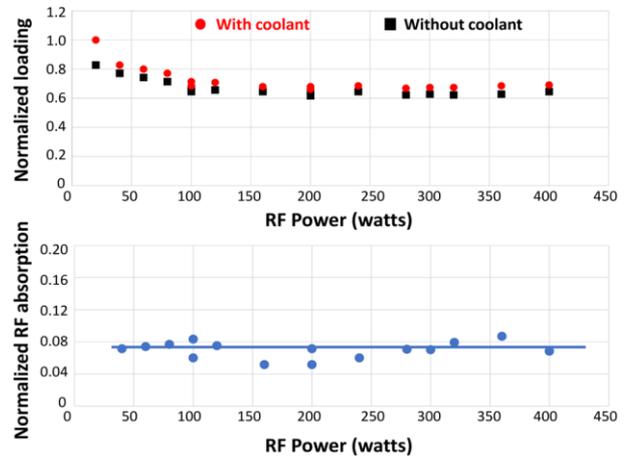

Fig. 5. Normalized vacuum loading with (red circles) and without (black squares) the DI water coolant in the source, and the corresponding RF absorption. The multiple points plotted for 100 W and 200 W are taken on two different days to show reproducibility. Please note that these experiments were done without plasma and hence for safety reasons, RF power was kept to less than 400 W.

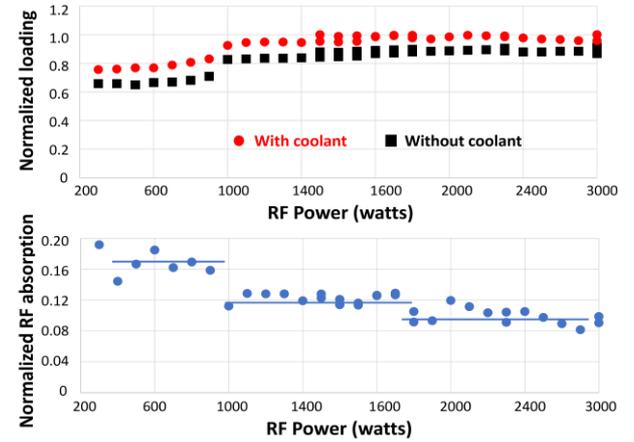

Fig. 6. Normalized loading with (red circles) and without (black squares) the DI water coolant in the source, and the corresponding RF absorption during plasma operation. The multiple points plotted for the same RF powers represent data taken on different days. These experiments were done with argon plasma at 50 sccm, 4 mTorr and 0.1 T of magnetic field.

As a final test of whether the DI water coolant affects high density plasma production, we show the radial profiles of the plasma density and the electron temperature with and without the DI water coolant, as measured by the RF compensated, voltage swept Langmuir probe in Fig(s). 7 and 8 respectively, obtained ~ 0.8 m downstream of the RF antenna, in argon plasma. For 2 kW of RF power, with ~ 4 mTorr (~ 0.53 Pa) of neutral argon gas flowing at 50 sccm and with a uniform magnetic field of 0.1 T, we achieve a peak electron density of ~ 1.95 x $10^{19}$ $m^{-3}$. The shots are very repeatable. The effect of the DI water coolant in the RF source on plasma production is negligible. The very slight differences can probably be attributed to the different matching conditions in the two cases, as also found in the VNA results. Introduction of the DI water coolant changes the effective RF antenna impedance and hence can slightly shift the resonance matching conditions, which can easily be taken care of by manual or automatic matching networks with variable capacitors. However, it is clear that the presence of the DI water coolant does not hinder high density plasma production.

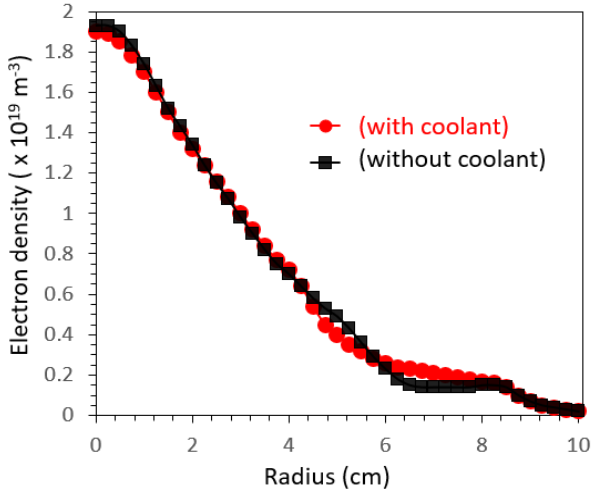

Fig. 7. Radial profiles of the plasma density, at 4 kW in argon, with (red circles) and without (black squares) the DI water coolant in the source. The chamber wall is at r = 10 cm.

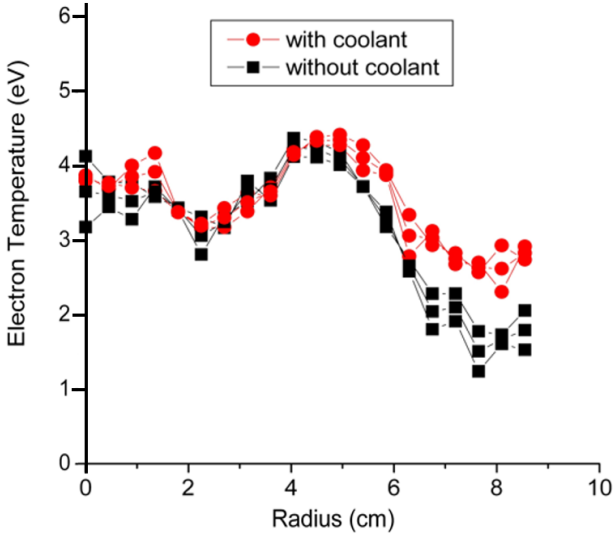

Fig. 8. Electron temperature, at 4 kW in argon, with (red circles) and without (black squares) the DI water coolant in the source. The scatter in the plots are from experiments done on 3 different days. The chamber wall is at r = 10 cm.

In a companion paper [42], we show systematic scans of argon plasma density, electron temperature and ion temperature as measured by laser induced fluorescence [LIF], as well as numerically computed profiles of the argon neutral density and temperatures. We show that the RF source design is stable enough to allow us to use LIF, which needed several minutes of integration of the emitted light for good signal to noise ratios to measure the ion velocity distribution functions of argon. Having confirmed this crucial step, henceforth we always performed experiments in the PISCES-RF with the DI water coolant flowing through the ceramic cylinders. This allowed us to safely increase the RF power to the helicon antenna to produce steady state plasmas at higher powers and perform thermal studies of the source.

## 4.2 Plasma heating and steady state temperature distributions on the plasma facing ceramic surface

We used IR imaging to investigate the thermal issues of the plasma facing inner ceramic surface, for steady state operations of the helicon source at RF powers of up to 20 kW. Strategically placed IR cameras allowed measurements of the spatio-temporal variation of the temperatures of the inner plasma facing ceramic surface. The IR camera was used to obtain two important pieces of information. We could calculate the effective heat deposited by the plasma on the inner ceramic. When this heat from the plasma was effectively removed by the DI water coolant, the system came to a thermal equilibrium within some characteristic time scale. This gave the final steady state temperature distribution (in the axial and azimuthal directions) of the inner ceramic surface.

The temperature, at each pixel, is measured as a function of time: $T(t)$. To calculate the effective heat deposited on the inner ceramic surface, we used the theory of heat conduction through a semi-infinite solid [43, 44]. The plasma turned on in less than a second and acted as a source of a constant heat flux on any point on the inner ceramic surface. Near the beginning of this constant heat flux, when the heat front hasn't yet had the time to reach the outer cooled surface of the ceramic, the approximation of a semi-infinite solid is valid. Since bodies of finite thickness can be treated as infinitely wide at the beginning of a constant heat pulse, the analytical solutions for this case could be applied (see section 2.3.3.1 of Ref. 43). In that approximation, the temperature, at each point on the inner ceramic surface, is given by

$$T(t) = T_0 + \left(\frac{2}{\sqrt{\pi}}\right) \cdot \left(\frac{\dot{Q}_0}{b}\right) \cdot \sqrt{t} \qquad (4)$$

where $T(t)$ is the instantaneous temperature, $T_0$ is the initial temperature, $\dot{Q}_0$ is the constant heat flux and $b$ is a constant which is determined by the thermal properties of the ceramic itself. Denoting $\rho$ as the density, $\lambda$ as the thermal conductivity and $C_p$ as the specific heat of the material at constant pressure, we have $b^2 = \rho \lambda C_p$. Rearranging and squaring equation (4), we get

$$[T(t) - T_0]^2 = K \cdot t \qquad (5)$$

which is effectively a straight line with slope $K$. Thus, from the experimentally measured slope $K$ and using the known material constants to calculate $b$, we could infer the steady state, constant heat flux on each point (experimentally, for each pixel on the IR camera screen that corresponds to a point on the ceramic) of the inner plasma facing ceramic surface from the expression:

$$\dot{Q}_0 = \left(\frac{\sqrt{\pi}}{2}\right) \cdot b \cdot \sqrt{K}. \qquad (6)$$

Finally, taking such inferred values of the steady state, constant heat flux from each pixel of each of the IR camera positions, we could construct the $z - \theta$ maps of the heat fluxes. In Fig. 9, we show the steady state, constant heat flux for 10 kW of RF power in argon plasma. We also show the footprint of the $m = 1$ RF antenna when unraveled and plotted on the $z - \theta$ plane. It is clearly seen that the largest heat deposition is under the straps of the helicon antenna, connected to the live, non-grounded, side of the RF power supply. This steady state, constant heat flux distribution occurs even though the plasma directly contacted the middle of the ceramic, near z = 0 (as the plasma would follow the slight expansion in the magnetic field flux lines due to the extra separation of the magnets near the RF antenna, as shown in Fig. 2). When integrated over the total surface of the cylinder, the total steady state, constant heat deposited is ~ 1.68 kW (16.8 % of the input RF power).

the ceramic are not directly in contact with the DI water coolant. We find that the regions under the helicon antenna strap remain colder even though they have the maximum heat deposited under them, thus signifying the effectiveness of the DI water coolant in removing the heat. Moreover, as mentioned earlier, the separation between the two magnets (as shown in Fig. 2) near the helicon antenna allows the magnetic field lines under the ceramic to bulge out. Thus, the plasma hits the ceramic at z = 0 which we envision to be the worst-case scenario for the MPEX source. Even then, we see that the maximum heating is under the straps of the helicon antenna and not at the spatial location where the plasma is supposed to hit the inner surface of the ceramic.

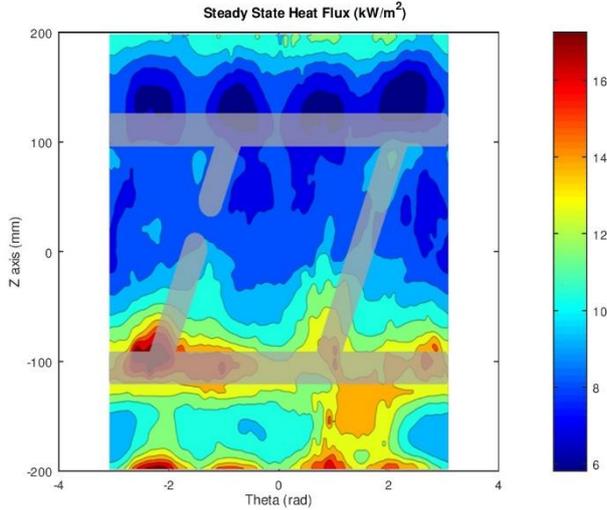

Fig. 9. The inferred 10 kW argon plasma heat flux on the plasma facing inner ceramic surface, for B = 0.1 T, along with the footprint of the helicon antenna superimposed (in semi-transparent, light gray).

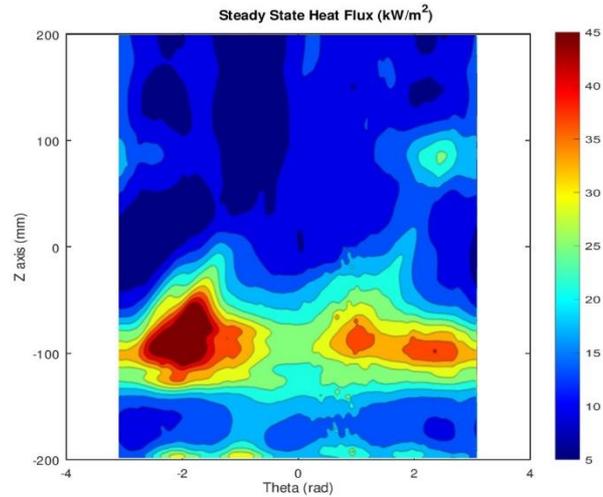

Fig. 11. The inferred 10 kW hydrogen plasma heat flux on the plasma facing inner ceramic surface, for B = 0.02 T.

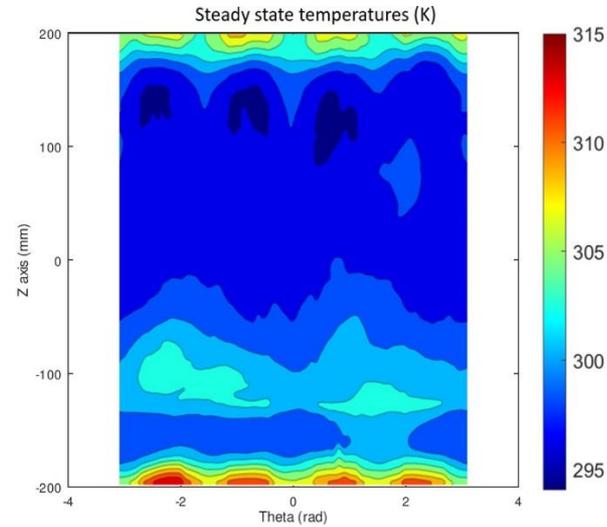

Fig. 10. Steady state temperature distribution due to 10 kW argon plasma at B = 0.1 T.

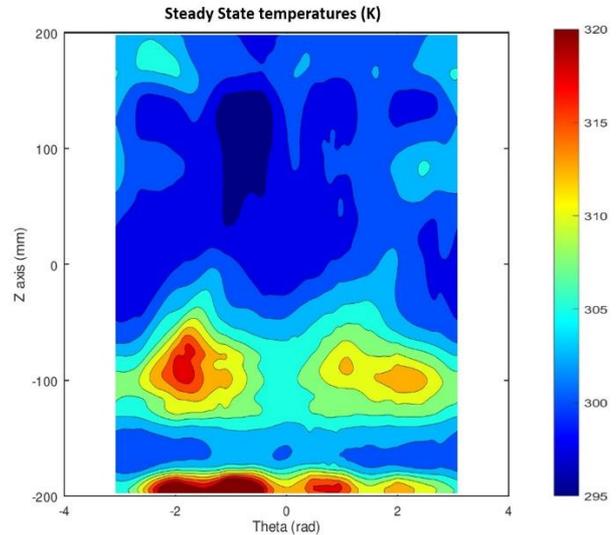

Fig. 12. Steady state temperature distribution, for 10 kW of hydrogen plasma at B = 0.02 T.

In Fig. 10, we show the spatial pattern of the steady state temperatures on the plasma facing inner surface of the ceramic, for 10 kW argon plasma and a DI water coolant flow rate of ~ 22 GPM (0.0014 m$^{-3}$s$^{-1}$). Before the plasma was turned on, the ceramic was kept at a uniform surface temperature of 293 K. The DI water coolant can remove most of the heat as the ceramic surface does not get very hot. The hottest portions of the ceramic are at the very ends. This is due to the constraints of the ceramic-metal brazing technique where ~ 1.5 cm of the edges of

Fig. 11 shows the final $z - \theta$ map of the steady state, constant heat flux for 10 kW of hydrogen plasma. The magnetic field profile is the same as used in argon, but the magnitude is much lower (0.02 T) in order to ensure proper RF matching. Again, it is clearly seen that the largest heat deposition is under the straps of the helicon antenna, connected to the live (non-grounded) side of the RF power supply. When integrated over the total surface area of the cylinder, the total heat deposited is ~ 2.76 kW, (27.6% of the input RF power). Thus, we find that the net heat deposited from the plasma is higher in case of hydrogen plasma when compared to argon plasma.

In Fig. 12, we see the spatial pattern of the steady state temperatures on the inner surface of the ceramic for 10 kW of steady state hydrogen plasma and a DI water coolant flow rate of ~ 22 GPM (0.0014 m$^{-3}$s$^{-1}$). The ceramic was at a uniform surface temperature of 293 K. We find that even though there is considerable heat deposition right under the helicon antenna straps (see Fig. 11), the DI water coolant is effective in cooling that area, but the hottest portions of the ceramic are near the very ends, which are not in direct contact with the DI water coolant, due to mechanical constraints of the titanium brazing. Even then, for 10 kW of RF power at steady state conditions, the maximum increase in the ceramic temperature is ~ 30 K, which would not produce catastrophic thermal stresses in the ceramic.

### 4.3 Calorimetry of the DI water coolant

To measure the net heat being removed by the DI water coolant, we used two thermocouples to measure the inlet and the outlet temperatures of the DI water. Knowing the flow rate of the DI water coolant, and measuring the temperature difference, we have the net heat (Q), given by

$$Q = \dot{m}.C.\Delta T \qquad (7),$$

where $\dot{m}$ is the mass flow rate of DI water, $C$ is the specific heat capacity of DI water and $\Delta T$ is the measured temperature difference. Since we had to take the difference of two large numbers to get a very small value, these measurements were prone to many errors. The thermocouples are extremely sensitive to RF noise, hence we had to ensure proper RF shielding of the cables with proper grounding to reduce RF pickup.

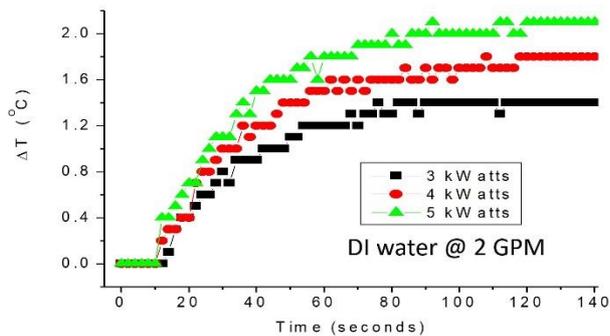

Fig. 13. Temperature difference between the outlet and the inlet of the coolant as a function of time.

For relatively low RF powers, if we used the maximum flow that was accessible to us using the recirculating chiller [~ 22 GPM (0.0014 m$^{-3}$s$^{-1}$)], the increase in the DI water temperature was of the same order of the noise. Hence, we added another valve to reduce the flow to the lowest possible value [~ 2 GPM (0.000126 m$^{-3}$s$^{-1}$)], so that the measured temperature differential was above the noise. Fig. 13 gives an example of such measurements, for relatively low RF powers, in high density argon helicon plasma. We see that as we increase the RF power, the steady state saturated value of the temperature differential also increases, which is consistent with expectations. Moreover, we also get an estimate of the time that the system takes to come to thermal equilibrium. For 2 GPM (0.000126 m$^{-3}$s$^{-1}$) of DI water flow, it takes ~ 2 minutes to reach the steady state. Using the data in Fig. 13 and equation (7), we find that the net heat removed by the DI water coolant for 3 kW, 4 kW and 5 kW of input RF power, is 756 W (25.2%), 972 W (24.3%) and 1134 W (22.7 %) respectively.

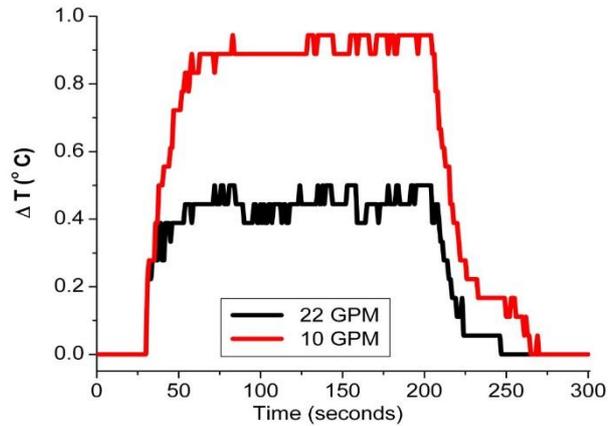

Fig. 14. Temperature difference between the outlet and the inlet of the coolant (DI water) as a function of time, at 10 kW of input RF power for argon plasma, shown for flow rates of 10 GPM (0.00063 m$^{-3}$s$^{-1}$) and 22 GPM (0.0014 m$^{-3}$s$^{-1}$), showing consistency of the calorimetry method.

In Fig. 14, we show another example of thermocouple data, but for higher powers in argon plasma. For 10 kW of input RF power the DI water coolant had to remove a much larger amount of heat, and hence we could obtain a higher signal to noise ratio even at higher flow rates of the DI water coolant. Here we also show the consistency in the net heat removed by changing the flow rates of the DI water coolant for the same RF power plasma. For 10 kW of argon plasma operation, calorimetry at 10 GPM (0.00063 m$^{-3}$s$^{-1}$) yields a net heat removal of 2.48 kW (24.8%) and for a flow rate of 22 GPM (0.0014 m$^{-3}$s$^{-1}$), yields 2.64 kW (26.4%), which are pretty consistent. We also notice that as we increase the flow rate, the time it takes to reach steady state levels decrease, as expected. For ~ 10 GPM (0.00063 m$^{-3}$s$^{-1}$), it takes ~ 1 minute and for ~ 22 GPM (0.0014 m$^{-3}$s$^{-1}$), which was the maximum flow rate for the recirculating chiller we are using, it takes ~ 30 secs to reach saturation. Moreover, even for steady state conditions, the increase in the temperature of the DI water coolant is nominal (< 1 K) which confirms the effectiveness of the cooling method.

### 4.4 High density hydrogen plasma production

Having confirmed high density argon plasma production (please see Fig. 7) while ensuring non-catastrophic thermal loading of the ceramic plasma facing component by efficient cooling, we increased the RF power to 20 kW to obtain steady state, high density hydrogen plasmas for achieving fluences relevant to PMI. For general safety purposes (since the PISCES-RF chamber was not water-cooled), we had the device on for ~ 2 minutes when operating at 20 kW. We ensured from calorimetry that the system reached steady state by then and also allowed us to finish recording all the data.

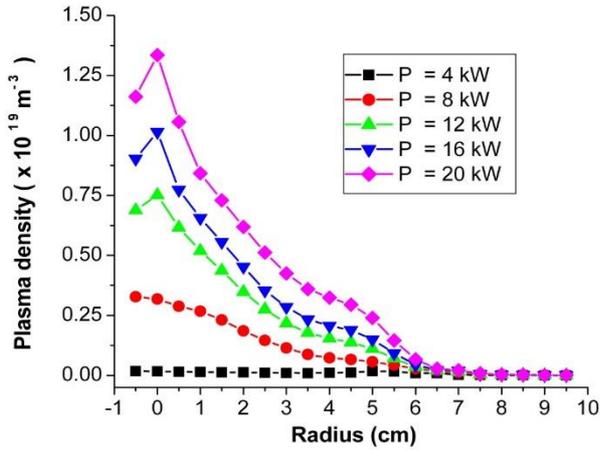

Fig. 15. Radial profiles of hydrogen plasma density measured 0.8 m downstream of the RF source, for B = 0.02 T.

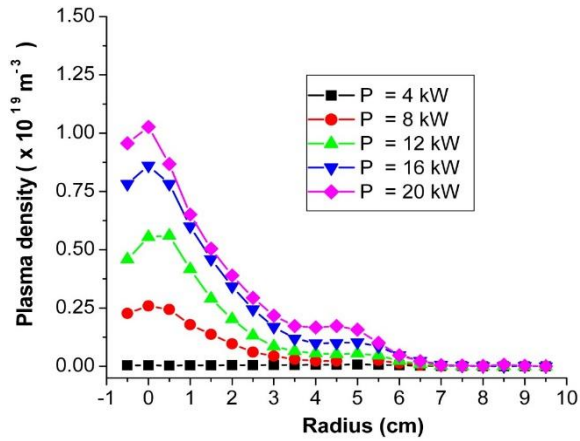

Fig. 16. Radial profiles of hydrogen plasma density measured 1.5 m downstream of the RF source, for B = 0.02 T. Please note that the ranges of the axes are kept identical to that of Fig. 16.

In Fig. 15 and Fig. 16, we show the radial profiles of the plasma density for steady state hydrogen plasmas as a function of RF power, measured 0.8 m and 1.5 m downstream of the RF helicon source, respectively. The source was operated with ~ 6 mTorr (~ 0.8 Pa) of neutral hydrogen gas flowing at 100 sccm and with uniform magnetic field of 0.02 T. Helicon sources have windows of operation in the operating parameter space (flow, pressure, magnetic field and RF power), and this above combination allowed us to achieve the helicon mode (W). We see from the plots that the helicon mode is achieved for RF power > 8 kW in our device. The plasma density at $r = 0$ is ~ $10^{17}$ m$^{-3}$ for up to 7 kW and jumps to ~ 0.3 x $10^{19}$ m$^{-3}$ at 8 kW and increases thereafter. A sudden discrete jump of more than an order of magnitude in the plasma density, as a function of RF power along with centrally peaked density profiles, is the standard signature of the helicon mode [12 – 17]. For 20 kW in hydrogen, we achieve peak central plasma densities of ~ 1.35 x $10^{19}$ m$^{-3}$ at the port 0.8 m downstream of the helicon source. From Fig. 16, we see that even 1.5 m downstream of the plasma source, the peak central densities are ~ 1 x $10^{19}$ m$^{-3}$, which makes this device PMI relevant. Moreover, even at 20 kW of steady state operation, the increase in the DI water temperature is only slightly more than 1 K, with the DI water coolant flow rate being 22 GPM (0.0014 m$^{-3}$s$^{-1}$).

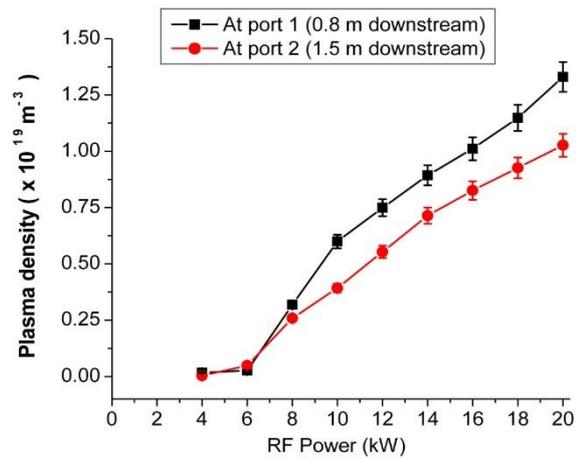

Fig. 17. Hydrogen plasma density vs. RF power measured at 0.8 m and 1.5 m downstream of the RF source, for B = 0.02 T.

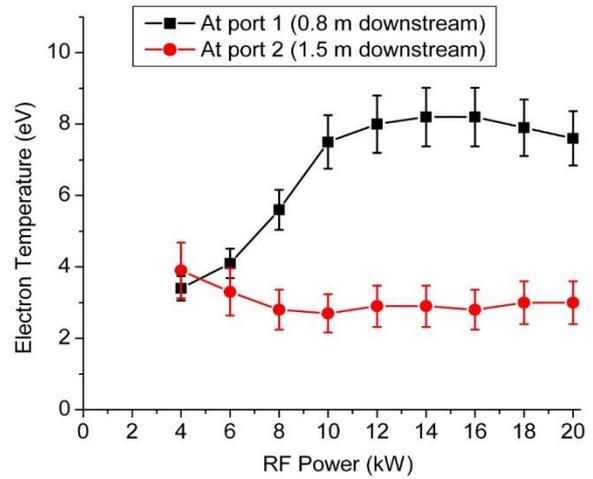

Fig. 18. Electron temperature vs. RF power measured at 0.8 m and 1.5 m downstream of the RF helicon source for hydrogen plasma, for B = 0.02 T.

In Figs. 17 and 18, we show the effect of increasing RF power on the central hydrogen plasma densities and electron temperatures measured at the two ports, 0.8 m and 1.5 m downstream of the source. After we achieve the helicon core mode for hydrogen plasma, in the range of the RF powers used in this study (up to 20 kW), the plasma density is linear in power and has not yet reached saturation. On the other hand, the electron temperature seems to have reached saturation (~ 8 eV at 0.8 m and ~ 3 eV about 1.5 m away from the RF source) once we have reached the high-density helicon core mode. From the measured ion saturation current, we calculate that the ion fluxes to the probes are ~ 3 x $10^{23}$ m$^{-2}$sec$^{-1}$ at 0.8 m and ~ 2 x $10^{23}$ m$^{-2}$sec$^{-1}$ at 1.5 m away from the RF source respectively. Note that the data shown (in Figs. 15 – 18) are taken for one particular set of plasma operating conditions: with ~ 6 mTorr (~ 0.8 Pa) of neutral hydrogen gas flowing at 100 sccm and with uniform magnetic field of 0.02 T. Plasma source parameter optimization to achieve the highest plasma densities or highest electron temperatures have not yet been done.

## 5. Conclusion and additional discussions

We have designed and built an actively cooled helicon plasma source with a DI water-cooled window for steady state plasma operation and also to test the possibility of using it in the 200 kW MPEX linear plasma device [23 – 25]. A prototype has been tested in the steady state up to 20 kW at UCSD. We showed data from high density plasma production in argon and hydrogen, by achieving the helicon plasma mode of operation (W), even with the DI water coolant flowing between the ceramic layers. At 20 kW of RF power, we can achieve hydrogen plasma density $> 10^{19}$ m$^{-3}$ and ion flux of $\sim 2 \times 10^{23}$ m$^{-2}$sec$^{-1}$ even 1.5 m away from the RF source. For these conditions, in the steady state, the increase in the DI water coolant temperature is $\sim 1$ K, which is nominal. It took $\sim 30$ secs for the DI water coolant to come to thermal equilibrium.

Simultaneously, we have studied different aspects of this source including RF absorption by the DI water coolant (see Figs. 4, 5 and 6), effect of the DI water coolant on high density plasma production (see Figs. 7 and 8), spatiotemporal measurements of heat deposition by the plasma and steady state temperature distributions on the inner plasma facing ceramic surface (see Figs. 9, 10, 11 and 12), the net heat that has to be removed for safe long-term operation of the device (see Figs. 13 and 14) and finally obtained high density hydrogen plasmas (see Figs. 15, 16, 17 and 18) with PMI relevant fluxes. Very recently, we also obtained the high-density helicon mode of operation with helium plasmas. We find that the windows of operation for helium plasmas are broader than that for hydrogen. For helium plasmas, we obtain the helicon mode at 3 kW for $\sim 9$ mTorr ($\sim 1.2$ Pa) of gas flow at 100 sccm and 0.045 T of external magnetic field. In the helicon mode of operation for helium plasmas, at $r = 0$ cm, we obtain $\sim 0.6 \times 10^{19}$ m$^{-3}$ of helium plasma density at 3 kW, which increases to $\sim 1.2 \times 10^{19}$ m$^{-3}$ for 20 kW. Peaked radial profiles, similar to that of argon and hydrogen confirm that we are in the helicon mode.

While achieving our goal of producing high density plasmas in steady state, we also learnt several things along the way. We have identified two possible physical mechanisms contributing to the net heating of the DI water coolant: plasma-induced *surface* heating of the dielectric window and *volumetric* heating due to RF absorption by the DI water coolant.

We found that for argon plasmas, $\sim 16 – 19\%$ of the net RF input power gets deposited on the ceramic layer directly due to the plasma hitting the inner ceramic (*surface* heating of the dielectric window). In addition, studies of the effect of the DI water coolant on the RF matching network with a VNA and Pearson transformer have shown that $\sim 8 – 11\%$ of the RF power is directly absorbed by the DI water coolant (*volumetric* RF heating of the coolant). Now, from calorimetry, using independent measurements of the inlet and outlet DI water temperatures, we find that $\sim 25 – 30\%$ of the input RF heat load is removed by the DI water coolant. Thus, the three independent thermal measurements of the different aspects of the antenna assembly seems to be consistent with each other, across a wide range of RF power. We represent them in Fig. 19. The black squares represent the RF absorption by the DI water in between the two ceramic cylinders. The red circles are calculated from the IR camera data, using the equations 4 – 6, which represent the plasma heat deposition on the inner ceramic, which in turn is transferred to the DI water coolant. The sum of the two sources of heat input to the DI water coolant is shown in magenta inverted triangles. We plot the measured heat removed by the DI water coolant from calorimetry, a completely independent measurement, as blue triangles. Within measurement errors, they are consistent over the range of 1.5 – 10 kW for argon plasma. Straight lines, going through the origin, fit to the data in Fig. 19, give slopes of 0.18 for the heat deposited on the plasma facing ceramic (as measured by IR imaging), 0.09 for the RF absorption (as measured by VNA) and 0.26 for the heat removed by the DI water (as measured by calorimetry).

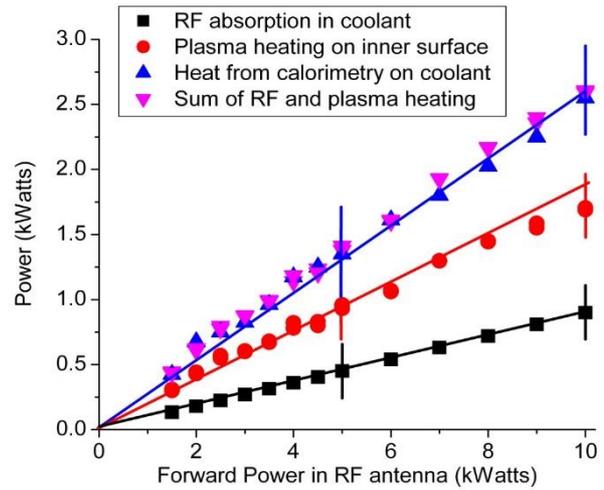

Fig. 19. Thermal trends with RF power and consistency among the different heating mechanisms for argon plasmas.

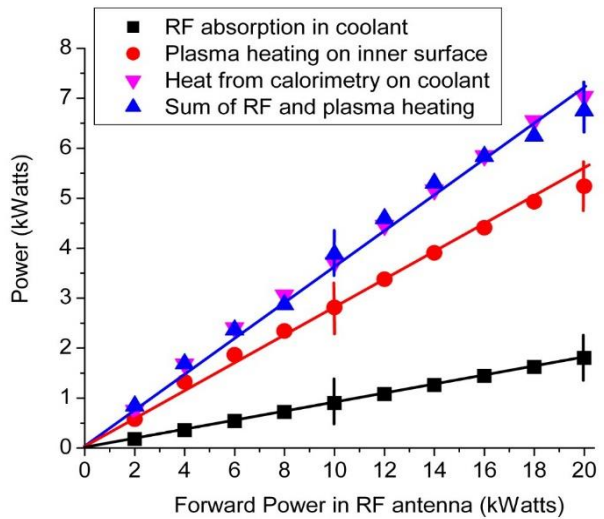

Fig. 20. Thermal trends with RF power and consistency among the different heating mechanisms for hydrogen plasmas.

Similarly, for hydrogen plasma, we find that the heat deposition by the plasma on to the inner ceramic surface is 25 – 29% and from calorimetry the net heat removal is 35 – 40 %. Taking the RF absorption to be 8 – 11 %, we find consistency for hydrogen plasmas too. The data is plotted in the Fig. 20 with straight line fits, going through the origin, giving slopes of 0.28 for the heat deposited on the plasma facing ceramic (as measured by IR imaging), 0.09 for the RF absorption (as measured by VNA) and 0.36 for the heat removed by the DI water (as measured by calorimetry). Recent work done on the device Proto-MPEX at ORNL shows ~ 30 – 40% of the input RF power being deposited on the RF transparent window [44, 48], which is also consistent with our finding.

In addition, we note that helicon sources have very strict windows of operation, in terms of the source parameters such as external magnetic field profile and magnitude near the antenna, gas flow, neutral gas pressure, RF power etc. These operating windows also depend on the gas being used and are very narrow for hydrogen, which is the most difficult gas to work with in terms of RF plasma generation in the helicon mode [45 – 47]. Optimization of the external source parameters for the highest possible densities and investigating the windows of operation for various fusion relevant gases such as helium, deuterium and hydrogen remain a task for the near future. Moreover, in the other examples of using RF helicon sources to generate high density hydrogen plasmas [18 – 21], the magnetic field near the target further downstream is kept much higher than the magnetic fields near the RF source (low magnetic fields needed for proper RF matching in hydrogen plasmas) and the mirror fields are used to increase the plasma density due to flux conservation. However, in this study, we have kept the magnetic field to be uniform in the downstream chamber. In that regard, we believe that the downstream plasma densities (near the target manipulator, which is supposed to be held at ~ 1.5 m downstream from the RF source in PISCES-RF) can be enhanced also by increasing the magnetic field near the target manipulator.

Moreover, for hydrogen plasmas, we find that the peak plasma density near the center seems to increase with increasing RF power, once it is in the helicon mode. This suggests that we can possibly increase the plasma density and ion fluxes further by going to higher RF powers. Typically, in a helicon plasma source, the plasma density can increase with increasing RF power, but the relation saturates at some value of RF power, and this is attributed to strong neutral depletion [49 – 52]. To understand the role of neutrals, we have used a Monte-Carlo based simulation [53] to estimate the radial profile of neutral density ($n_n$) at the two axial ports, 0.8 and 1.5 m away from the RF helicon source. The simulation is constrained by the gas pressure measured with a neutral pressure gauge. Experimentally measured plasma densities and the electron temperatures are used as input parameters. Ionization by electron impact, neutral-neutral, and ion-neutral collisions are included, but plasma and neutral flows and wall pumping are ignored. Using a 0-D hydrogen molecule simulation [54] balancing the rate coefficients for hydrogenic ion formation and destruction (the measured mean densities and electron temperatures are input parameters), we can get a rough idea of the ion concentrations. For hydrogen plasma at 8 kW, we have approximately 34.6 % of H+, 16.9 % of H2+ and 48.6 % of H3+ ions. For 20 kW of hydrogen plasma, we get approximately 57.6 % of H+, 23.7 % of H2+ and 18.7 % of H3+ ions. Basically, as an increasing function of RF power, we see less H3+, and more H2+ and H+ ions.

The neutrals are expected to be dominantly H2, while ions are expected to be mostly H3+ at the lower RF powers and dominated by H+ at the higher RF powers. Due to the complexity of the situation, the Monte-Carlo model tracks H2 molecules and assumes that the ions are H2+. In Fig. 21 and 22, we show the radial profiles of neutral densities. The details can change with source parameters, but the shapes remain similar, as long as we are in the high-density helicon mode of operation (W). We find strong radial neutral depletion near the core of the plasma, when we are in the high-density helicon mode (W), as we increase the RF power. This is consistent with previous work [49 – 52] in helicon sources with centrally peaked radial profiles of plasma density.

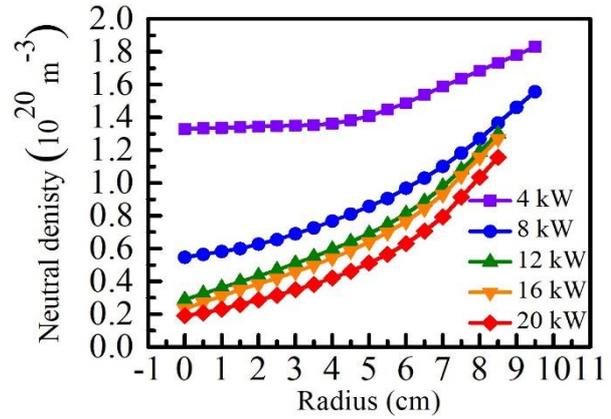

Fig 21. Radial profiles of the calculated neutral densities for hydrogen at 0.8 m downstream of the RF helicon source.

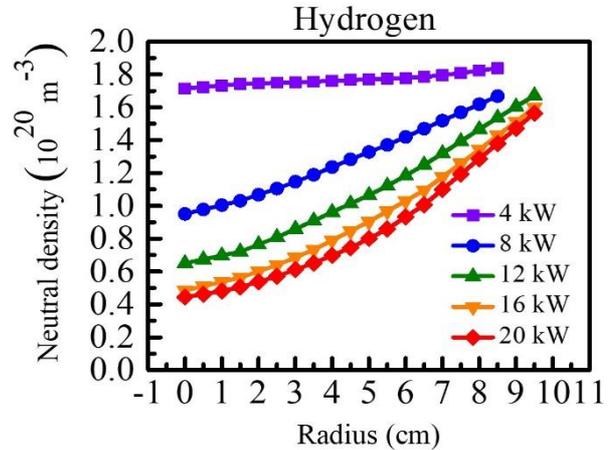

Fig 22. Radial profiles of the calculated neutral densities for hydrogen at 1.5 m downstream of the RF helicon source.

Moreover, in this experiment, the gas injection was upstream of the RF source region and as has been shown in [40], the high-density plasma that forms under the RF antenna acts like a choke for the neutrals. So, axially there is strong ionization, and hence neutral depletion, under the RF source as the gas flows from the upstream ($n_n$ ~ 2 x $10^{20}$ m$^{-3}$) of the RF antenna to the downstream ($n_n$ ~ 0.2 x $10^{20}$ m$^{-3}$ and $n_n$ ~ 0.4 x $10^{20}$ m$^{-3}$ measured at 0.8 m and 1.5 m away from the RF source). This is also consistent with previous studies of axial variation of neutrals in helicon sources [40, 52].

Finally, we would like to conclude by saying that in addition to developing and studying the DI water cooled plasma source for future PMI experiments in the PISCES-RF device, we are also using results from these experiments as a source design test for the 200 kW MPEX device, that is planned to be a world leading linear plasma facility to expose neutron irradiated samples to high ion fluences ($10^{31}$ m$^{-2}$) to examine multivariate PMI effects that will be present in next step fusion devices [23, 24]. For argon plasma, from the IR camera measurements, the net heat flux integrated over the whole inner ceramic, is given by a scaling law: $\dot{Q}_0 \sim 0.168 * RF\ Power$ (in kW) (see Fig 19). Similarly, for hydrogen helicon plasma, the scaling law would be given by $\dot{Q}_0 \sim 0.284 * RF\ Power$ (in kW) (see Fig 20). Understanding these trends are important when trying to extrapolate thermal test results from PISCES-RF to other devices being planned, such as the MPEX. These experimental results are being used as boundary conditions and initial conditions in thermal-hydraulic (Computational Fluid Dynamics: CFD) modelling and thermal-structural (Finite Elements: FE) simulations to answer critical engineering questions that arise in designing the 200 kW RF source for the MPEX device at ORNL [26].

## Acknowledgments

This research is sponsored by the Office of Fusion Energy Sciences, U.S. Department of Energy, under contract DE-FG02-07ER54912. The authors also want to thank Mr. Leopoldo Chousal and Mr. Tyler Lynch for their help in technical details on mechanical issues of the source design.